\documentclass[slac_one]{revtex4}
\usepackage{graphicx}
\usepackage{fancyhdr}
\begin{document}

\title{{\small{2005 ALCPG \& ILC Workshops - Snowmass,
U.S.A.}}\\ %% Please keep this conference title here
\vspace{12pt}
Distinguishing Between Models with Extra Gauge Bosons at the ILC} %% Paper title goes here


\author{Stephen Godfrey, Pat Kalyniak, Alexander Tomkins}
\affiliation{Ottawa Carleton Institute for Physics, 
Department of Physics, Carleton University, Ottawa K1S 5B6 Canada}

\begin{abstract}
Extra neutral gauge bosons arise in many extensions of the Standard Model.  
If one were to be discovered it would be necessary to measure its 
properties so that we could understand its origins. In this report we 
find that $Z'$ couplings can be measured at the ILC precisely enough to 
distinguish between models up to a $Z'$ mass of 2-3~TeV.  An important 
ingredient in these measurements is polarization of the $e^-$ and to a 
lesser extent $e^+$ beams. $b$ and $c$-quark tagging would also give 
important additional information. 
\end{abstract}

\maketitle

\thispagestyle{fancy}


\section{INTRODUCTION} % Section title should be in all capitals.

Extra neutral gauge bosons ($Z'$) and other $s$-channel resonances are 
predicted by many models of new physics. Previous studies have 
examined $Z'$'s predicted by string inspired models, LR symmetric 
models and numerous other extended gauge theories
\cite{Leike:1998wr,Cvetic:1995zs,Aguilar-Saavedra:2001rg,Weiglein:2004hn,Godfrey:1994qk,Capstick:1987uc}.  
Recently there has 
been a resurgence of interest in extra gauge bosons as they arise in 
many theories of current theoretical interest - in models with extra 
dimensions as Kaluza Klein excitations of the photon and $Z^0$
\cite{Rizzo:2004kr} and in 
the various manifestations of the Little Higgs Models 
\cite{Schmaltz:2004de,Arkani-Hamed:2002qy,Schmaltz:2005ky,Han:2003wu,Han:2005ru}.  
In many scenarios it is possible that they are 
low enough in mass to be discovered by the LHC.  
However, this is but the first step to determine the underlying 
theory.  To distinguish between the numerous possibilities it will be 
necessary to measure the properties of these new resonances. 

The International Linear Collider is ideally suited to this task as 
it can make precision measurements of 
various observables starting with the basic process 
$e^+e^-\to f\bar{f}$ where  $f$ 
could be leptons $(e,\; \mu ,\; \tau)$ or quarks $(u, \; d, \; c,\; 
s,\; b)$ (and possibly the $t$-quark).  This 
topic has been studied  in a number of places, most notably in the work 
of S. Riemann as reported in Ref.~\cite{Aguilar-Saavedra:2001rg}.  
In this contribution we expand on the work of Riemann by
including a wider set of models of new physics and by exploring the 
sensitivity of the results to electron and positron polarization, 
to the $Z'$ mass, and to including additional observables in the fits.  
In particular, we examine what can be learned by 
including $c$ and $b$-quark tagging to remove the ambiguities when 
only leptonic observables are included. In this contribution we 
concentrate on our new results on $Z'$ properties.  More generally
other forms of $s$-channel resonances appear in theories of current 
interest.  For example, theories of extra dimensions predict 
Kaluza Klein towers of massive gravitons. 
These are spin 2 objects so that they can be distinguished 
from $Z'$'s by measuring the angular distribution of their decay 
products, either directly at the LHC or indirectly at the ILC.  

We start with a short 
description of the observables with special emphasis on polarization. 
We then give our results for the various scenarios considered.  A more 
detailed report will be given elsewhere \cite{godfrey}.

\section{OBSERVABLES IN $e^+e^-$ COLLISIONS}

At $e^+e^-$ colliders, precision measurements see the effects of new 
$s$-channel resonances through deviations from standard model 
predictions due to interference between the $Z'$'s and the photon
and SM $Z^0$.
The basic process is $e^+e^- \to f\bar{f}$ where $f$ 
could be leptons $(e,\; \mu ,\; \tau)$ or quarks $(u, \; d, \; c,\; 
s,\; b)$.  From the basic reactions a number of observables can be 
used to search for the effects of $Z'$'s: The leptonic cross 
section, $\sigma (e^+ e^- \to \mu^+ 
\mu^-)$, the ratio of the hadronic to the QED point cross section,
$R^{had}= \sigma^{had}/\sigma_0$, the leptonic forward-backward 
asymmetry, $A^\ell_{FB}$, the leptonic  longitudinal asymmetry, 
$A^\ell_{LR}$, the hadronic longitudinal asymmetry, $A^{had}_{LR}$, 
the forward-backward asymmetry for specific quark or lepton 
flavours, $A^f_{FB}$, the $\tau$ polarization asymmetry, 
$A_{pol}^\tau$, and the polarized forward-backward asymmetry for 
specific fermion flavours, $A^f_{FB}(pol)$. The indices
indicate the final state fermions where $f=\ell$ and
$q$, with $\ell =(e,\mu,\tau)$, $q=(c, \; b)$, 
and $had=$`sum over all hadrons'.
The expressions for these observables are given in ref. 
\cite{Capstick:1987uc,e+e-}.  
A deviation for one observable is always possible as 
a statistical fluctuation so a more robust strategy is 
to combine many observables to obtain a $\chi^2$ figure of merit.  We 
follow this approach by including various observables.

\subsection{Polarization}
An important ingredient in precision measurements at the ILC is the use of 
electron and positron polarization 
\cite{Moortgat-Pick:2005cw,Fujii:1995ys}. These were included in the 
results given by Riemann in Ref.~\cite{Aguilar-Saavedra:2001rg}.   
%Riemann included polarization in her results although she did not give 
%the details of her analysis. 
Although these details are given in 
Ref.~\cite{Moortgat-Pick:2005cw,Fujii:1995ys} we reproduce the results 
here for the benefit of the interested reader.  The cross section at 
an $e^+e^-$ collider with longitudinally-polarized beams can be 
written as:
\begin{eqnarray}
\sigma_{P_{e^-} P_{e^+}} & =& \frac{1}{4} \left\{  
(1+P_{e^-})(1+P_{e^+})\sigma_{RR} + (1-P_{e^-})(1-P_{e^+})\sigma_{LL} 
\right. \cr 
& & \qquad \left. (1+P_{e^-})(1-P_{e^+})\sigma_{RL} + 
(1-P_{e^-})(1+P_{e^+})\sigma_{LR}  \right\}
\label{eqn1}
\end{eqnarray}
where $\sigma_{RL}$ stands for the cross section if the $e^-$ beam is 
completely right-handed polarized ($P_{e^-}=+1$) and the $e^+$ beam is 
completely left-handed polarized ($P_{e^+}=-1$) and analogously for
$\sigma_{LR}$, $\sigma_{RR}$, and $\sigma_{LL}$.  
For $s$-channel production of vector bosons only $\sigma_{LR}$ 
and $\sigma_{RL}$ contribute.  
For this case the cross section for arbitrary polarizations 
can be written as
\begin{equation}
\sigma_{P_{e^-}P_{e^+}} = (1-P_{e^+}P_{e^-}) \sigma_0[1-P_{eff}A_{LR}]
\end{equation}
where $\sigma_0=(\sigma_{RL}+\sigma_{LR})/4$
is the unpolarized cross section,
$A_{LR}=(\sigma_{LR}-\sigma_{RL})/(\sigma_{LR}+\sigma_{RL})$ is the 
left-right asymmetry, and $P_{eff}=(P_{e^-}-P_{e^+}) /(1-P_{e^+}P_{e^-})$ 
is the effective polarization. One sees that 
the collision cross sections can be 
enhanced if both beams are polarized and if $P_{e^-}$ and $P_{e^+}$ 
have different signs. This can be parametrized in an effective luminosity
given by
\begin{equation}
{\cal L}_{eff}= \frac{1}{2} (1-  P_{e^-}P_{e^+}) {\cal L}
\end{equation}
Thus, one can obtain a  $P_{eff}$ much higher than either of the two 
beam polarizations in addition to enhancements in ${\cal L}_{eff}$.
Another important result is that the uncertainty $\Delta P_{eff}/ P_{eff}$ 
is less than the uncertainty of the individual polarizations 
$\Delta P_{e^-}/ P_{e^-}$.  The improvement in the measurements due to 
positron beam polarization can be substantial.  One should see 
Ref.~\cite{Moortgat-Pick:2005cw,Fujii:1995ys} for a more complete 
discussion.

\section{RESULTS}

We are interested in answering the question of how well we can 
distinguish between $Z'$'s originating from different models.  We take 
as our starting point the analysis of Riemann which used 
leptonic observables to demonstrate that one can extract $Z'$ 
couplings and discriminate between models 
\cite{Aguilar-Saavedra:2001rg}.  
In this brief report we explore the sensitivity of her results to 
variations in the assumptions used in obtaining those results.
As mentioned in the introduction, numerous models exist.  For the 
purposes of this study we consider $Z'$'s coming from the $E_6$ $\chi$ 
model ($\chi$), 
LR-symmetric model (LR), Littlest Higgs model (LH)
\cite{Arkani-Hamed:2002qy,Han:2003wu}, 
Simplest Little Higgs model (SLH) \cite{Schmaltz:2004de},
and KK excitations originating in theories of extra dimensions (KK). 
We only present results for the Simplest Little Higgs model 
with a universal fermion sector. 
The KK case is problematic since, for this case,
the couplings shown do not in 
fact correspond to the KK $Z'$ couplings because in this model there are 
both photon and $Z^0$ KK excitations roughly degenerate in mass.  
The point is simply that the KK model can be distinguished from other models.

To obtain our plots, unless otherwise stated, we took 
$\sqrt{s}=500$~GeV and ${\cal L}_{int}=1$~ab$^{-1}$
assuming electron and positron polarization of 80\% and 60\% 
respectively, $\Delta P_{e^\pm}=0.5\% $, $\Delta {\cal L}=0.5\%$, and 
a systematic error of $\Delta ^{sys}=0.25\%$. The 
$Z'$ couplings shown in the figures are normalized such that the SM 
$Z^0$ couplings are 
$C^e_L=-{1\over 2}+\sin^2\theta_w$
and $C^e_R=\sin^2\theta_w$.

\begin{figure*}[t]
\centering
\includegraphics[width=75mm]{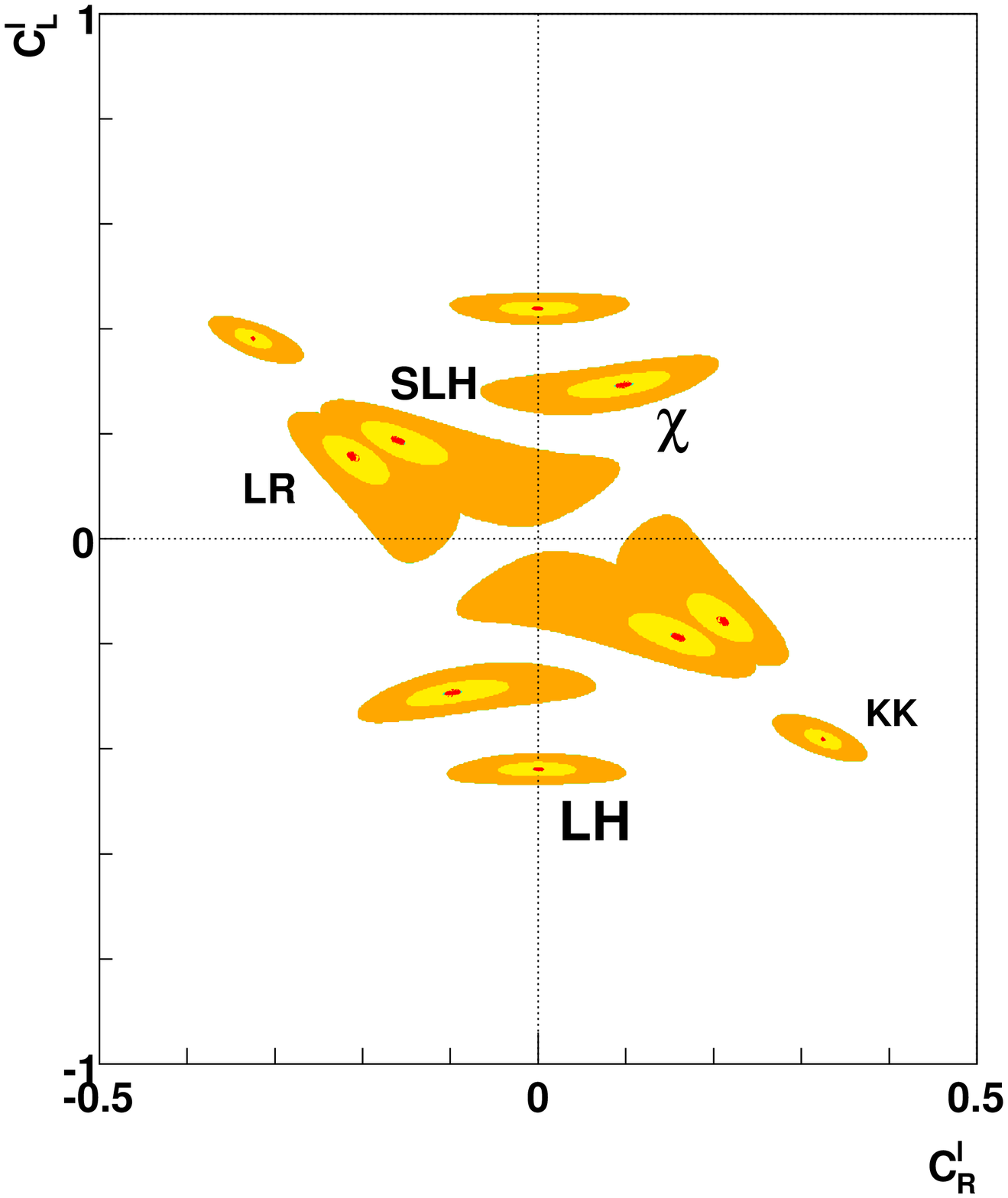}\qquad\qquad
\includegraphics[width=75mm]{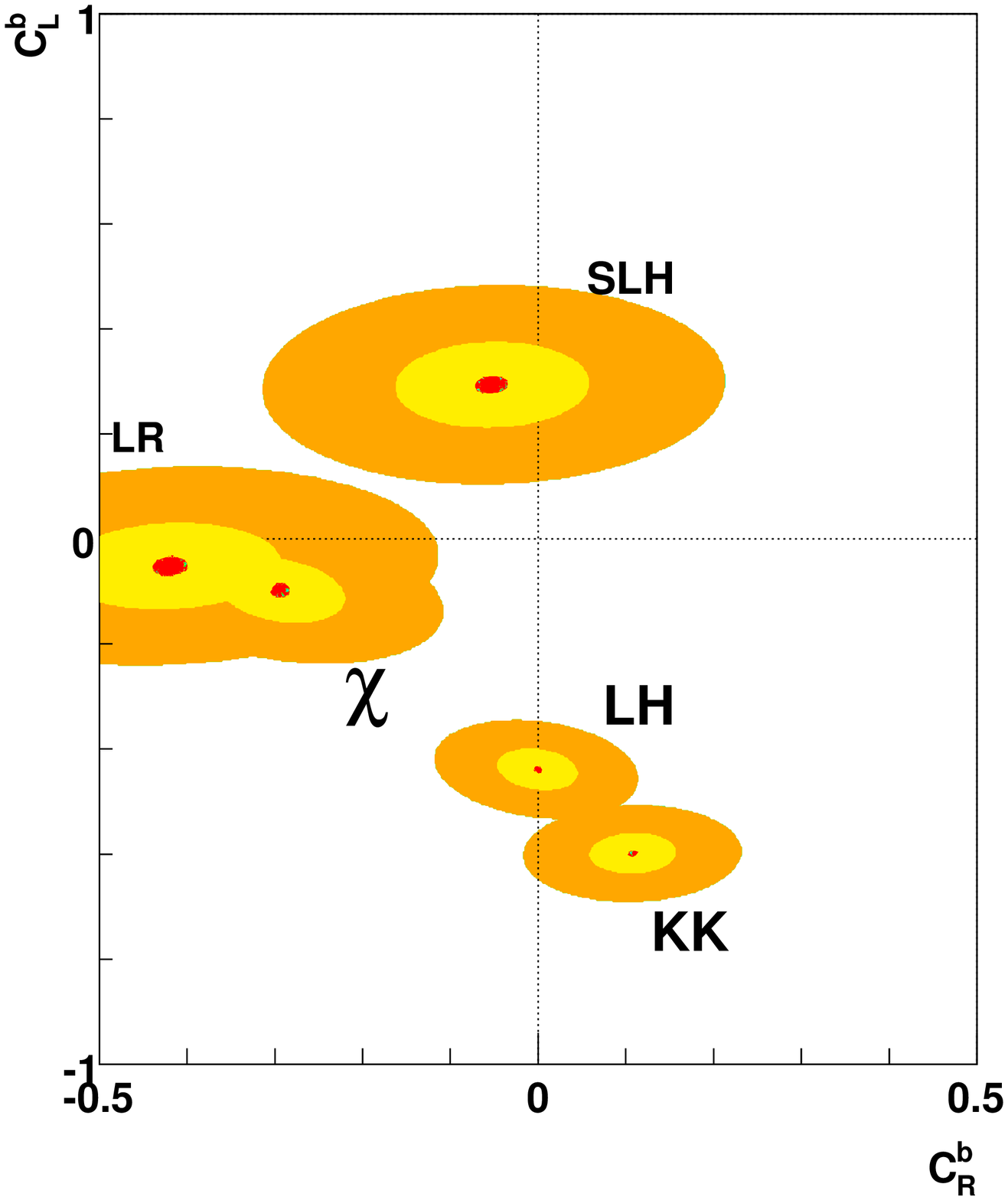}
\caption{Resolving power (95\% CL) for $M_{Z'}=1$, 2, and 3~TeV
and $\sqrt{s}=500$~GeV, ${\cal L}_{int}=1$ab$^{-1}$. The smallest 
regions correspond to $M_{Z'}=1$~TeV and the largest to $M_{Z'}=3$~TeV.
The left side is for leptonic couplings based on the leptonic 
observables $\sigma_{P_{e^-}P_{e^+}}^\mu$, $A_{LR}^\mu$, $A_{FB}^\mu$.  
The right side is for $b$ 
couplings based on the $b$ observables $\sigma_{P_{e^-}P_{e^+}}^b$, 
$A_{FB}^b$, $A_{FB}^b(pol)$
assuming that the leptonic couplings are known and a $b$-tagging 
efficiency of 70\%. 
%The couplings correspond to the $E_6$ $\chi$, LR, LH, SLH and KK models.  
} \label{fig1}
\end{figure*}

Fig.~\ref{fig1}(a) shows the resolving power of the lepton couplings 
assuming lepton universality and using the three observables: 
$\sigma_{P_{e^-}P_{e^+}}^\mu$,
$A_{FB}^\mu$ and $A_{LR}^\mu$ for $M_{Z'}= 1$, 2 and 3~TeV. 
As noted by Riemann there is a two-fold 
ambiguity in the signs of the lepton couplings since all lepton 
observables are bilinear products of the couplings. The hadronic 
observables can be used to resolve this ambiguity since for this case 
the quark and lepton couplings enter the interference terms 
linearly.  Fig.~\ref{fig1}(b) shows the resolving power for $b$-quark 
couplings based on the $b$-quark  observables $\sigma^b$, $A_{FB}^b$, 
$A_{FB}^b(pol)$
assuming that the leptonic couplings are accurately known from 
other measurements
and a $b$-tagging efficiency of 70\%.  One could gain additional 
information by studying other observables with hadron final states 
such as $R^{had}$, $A^{had}_{LR}$, and observables involving the $c$-quark.

We next consider the importance of polarization. 
In Fig.~2 we show results for the cases of no polarization, only 
the electron is polarized, and both the electron and positron are 
polarized.  The results are shown for $M_{Z'}=2$~TeV, 
$\sqrt{s}=500$~GeV and ${\cal L}_{int}=1$ab$^{-1}$
using the three observables  $\sigma_{P_{e^-}P_{e^+}}^\mu$,
$A_{LR}^\mu$, $A_{FB}^\mu$.  Note that the appropriate values of 
$P_{e^-}$ and $P_{e^+}$ are used in eqn.~\ref{eqn1} and for
the unpolarized case $A_{LR}$ does not contribute.  Clearly 
polarization will be important for measuring couplings and 
disentangling models if a $Z'$ were discovered although positron 
polarizaton does not appear to be an important factor for these 
measurements.

In Fig.~2 we assumed a $Z'$ mass of 2~TeV.  But the LHC has 
the potential of discovering a heavy neutral gauge boson up to 5~TeV 
or higher.  Supposing that this is the case, can the ILC still give us 
useful information?  In Fig.~3 we show the resolving power 
for $Z'$'s with $M_{Z'}=1$, 2, 3, and 4~TeV, 
again using only the three $\mu$ 
observables assuming the $e^-$ and $e^+$ polarizations given above.  
Reasonably good measurements can be made for 
the $M_{Z'}=2$~TeV case. For $M_{Z'}=3$~TeV the resolving power deteriorates 
but the measurements can still distinguish between many 
of the currently popular models.  At $M_{Z'}=4$~TeV it becomes quite 
difficult to distinguish among the models although some 
models could still be ruled out.

\begin{figure*}[t]
\begin{minipage}[t]{8.5cm}
\centering
\includegraphics[width=75mm]{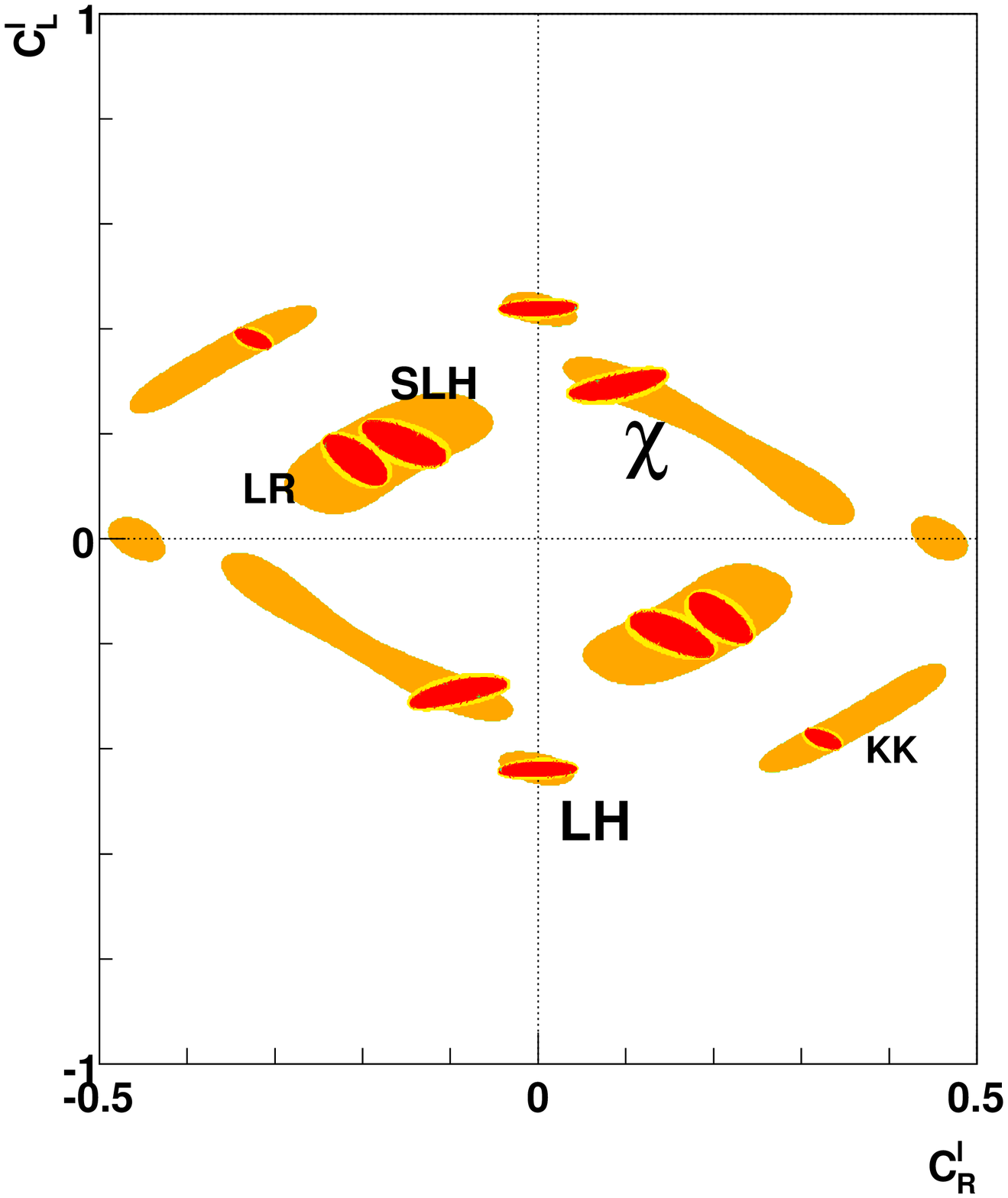}
\caption{The effect of polarization on coupling measurements.
Resolving power (95\% CL) for $M_{Z'}=2$~TeV
and $\sqrt{s}=500$~GeV, ${\cal L}_{int}=1$ab$^{-1}$ 
for leptonic couplings based on the leptonic 
observables 
$\sigma_{P_{e^-}P_{e^+}}^\mu$, $A_{LR}^\mu$, and  $A_{FB}^\mu$. The 
largest region corresponds to the unpolarized case while the smallest 
region corresponds to electron and positron polarization of
of 80\% and 60\% respectively with the middle region corresponding to 
only electron polarization.
%The couplings correspond to the $E_6$ $\chi$, LR, LH, SLH, and KK models. 
} 
\end{minipage} \qquad
\begin{minipage}[t]{8.5cm}
\centering
\includegraphics[width=75mm]{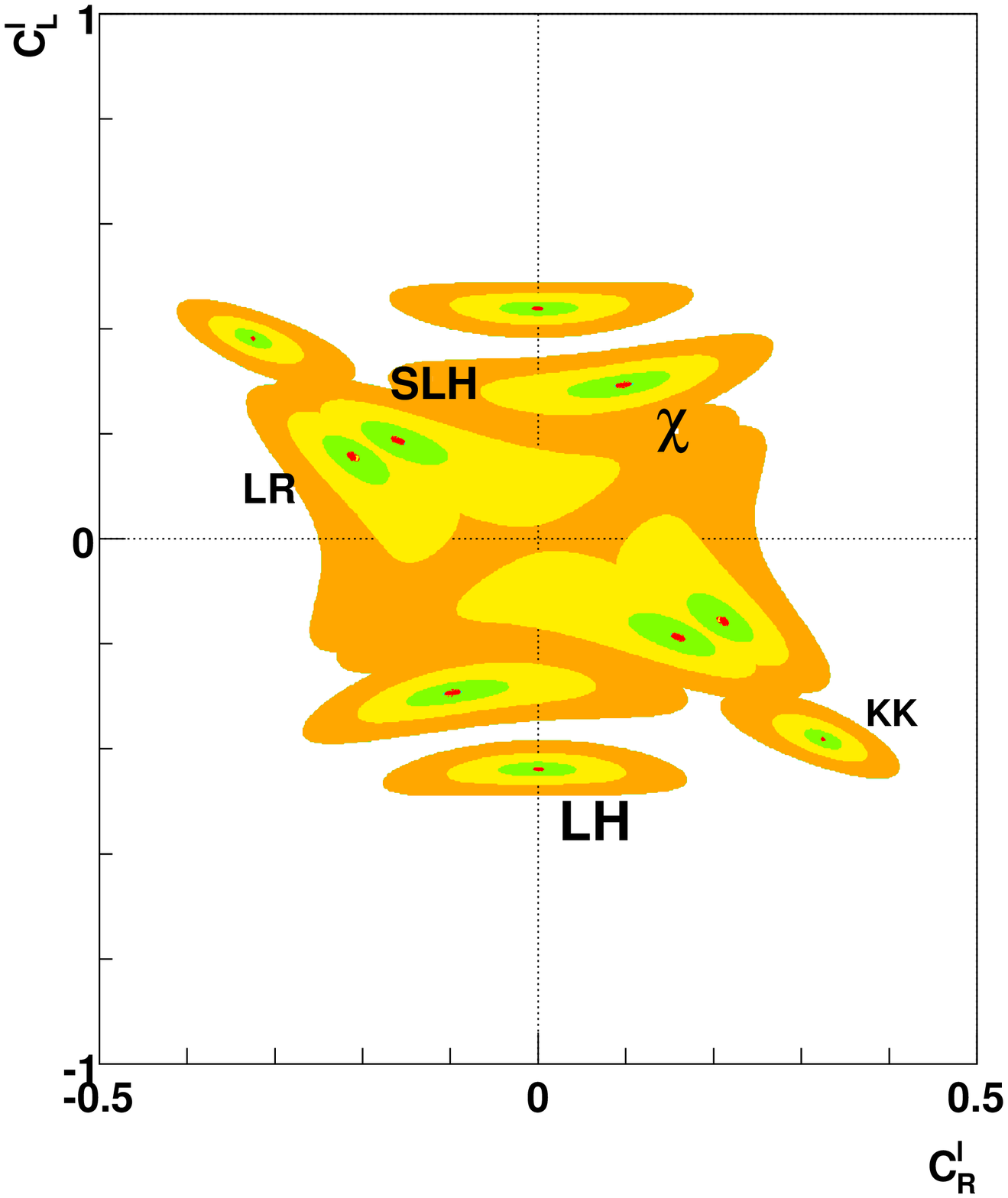}
\caption{Resolving power (95\% CL) for $M_{Z'}=1$, 2, 3, and 4~TeV,
and $\sqrt{s}=500$~GeV, ${\cal L}_{int}=1$ab$^{-1}$ 
for leptonic couplings based on the leptonic 
observables 
$\sigma_{P_{e^-}P_{e^+}}^\mu$, $A_{LR}^\mu$, and  $A_{FB}^\mu$. 
%The couplings correspond to the $E_6$ $\chi$, LR, LH, SLH, and KK models.
} 
\end{minipage} 
\end{figure*}

\begin{figure*}[t]
\centering
\includegraphics[width=75mm]{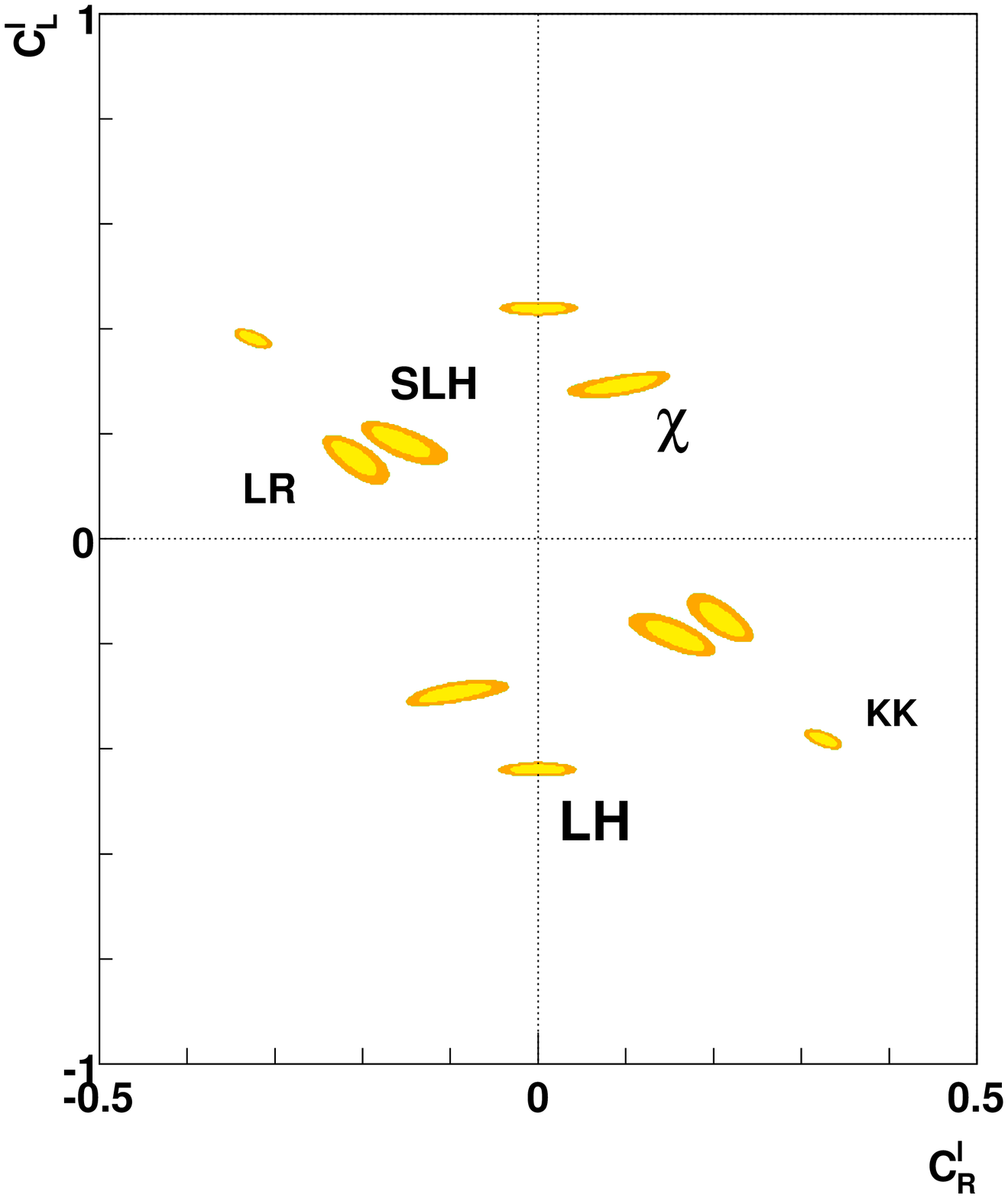}\qquad\qquad
\includegraphics[width=75mm]{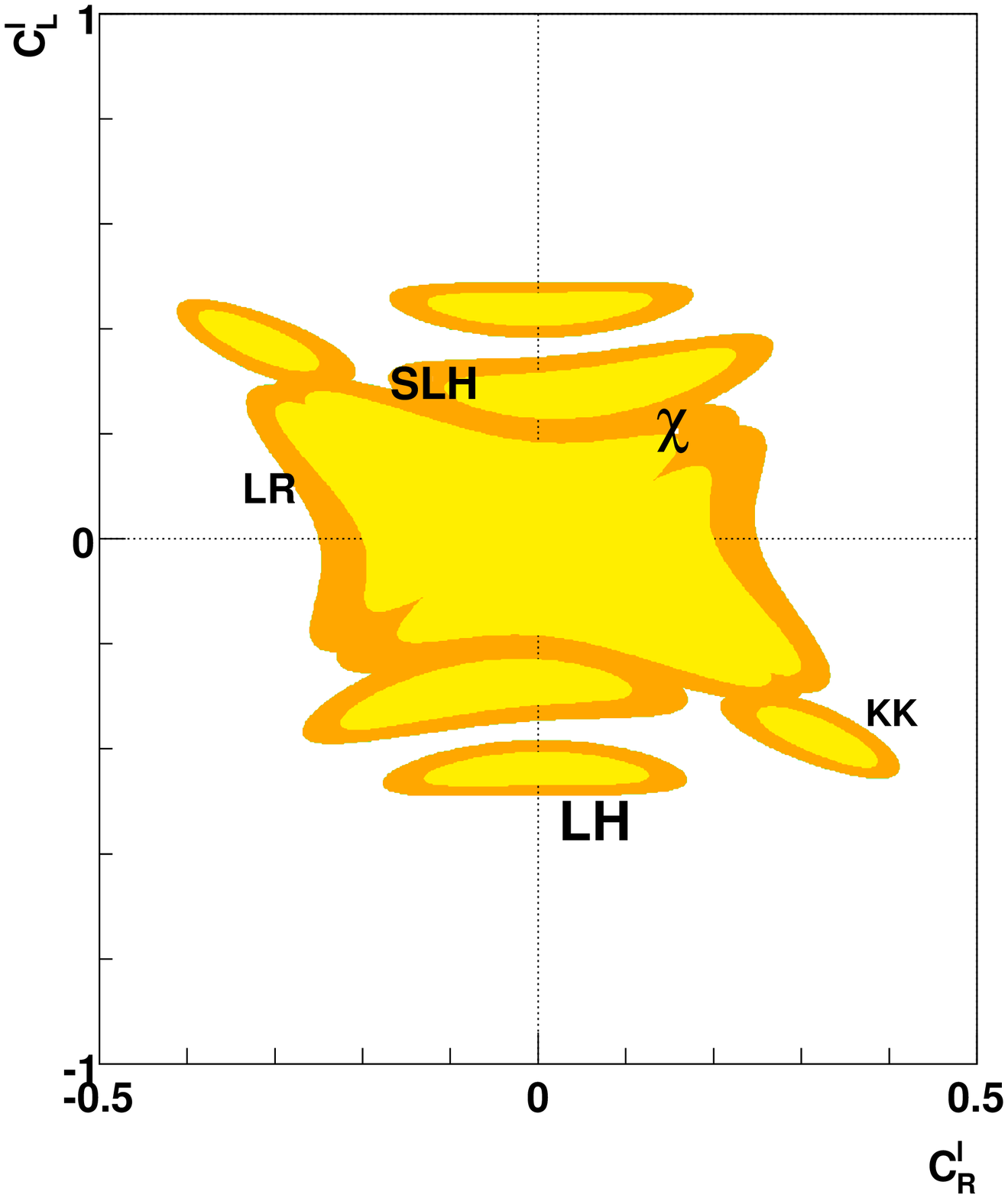}
\caption{Resolving power (95\% CL) of leptonic couplings 
for $M_{Z'}=2$~TeV (left side) and
$M_{Z'}=4$~TeV (right side)
and $\sqrt{s}=500$~GeV, ${\cal L}_{int}=1$ab$^{-1}$.
The outer region only includes the three muon observables
$\sigma_{P_{e^-}P_{e^+}}^\mu$, $A_{LR}^\mu$, and  $A_{FB}^\mu$ while 
the smaller region includes, in addition, the five tau  observables
($\sigma_{P_{e^-}P_{e^+}}^\tau$,  
$A_{LR}^\tau$, $A_{FB}^\tau$, $A_{FB}^\tau(pol)$, and $P_\tau$).
%The couplings correspond to the $E_6$ $\chi$, LR, LH, SLH, and KK models. 
%From Ref. \cite{godfrey}.
} \label{fig4}
\end{figure*}

In Fig.~\ref{fig4} we examine possible improvement in the resolving 
power by including more observables.  In the previous figures we only 
included three observables with final state muons.
If $\tau$ leptons could be observed 
with reasonable efficiency an additional five observables  
($\sigma_{P_{e^-}P_{e^+}}^\tau$,  
$A_{LR}^\tau$, $A_{FB}^\tau$, $P_\tau$ the $\tau$ polarization, and
$A_{FB}^\tau(Pol)$)
can be 
included in the $\chi^2$. Fig.~\ref{fig4} shows the improvement one 
gains by including the $\tau$ observables for $M_{Z'}=2$~TeV (left 
figure)  $M_{Z'}=4$~TeV (right figure).  For lack of a better estimate 
we simply take the $\tau$ efficiency equal to one which is clearly 
overly optimistic.  For the $M_{Z'}=$2~TeV case the improvement is not 
so impressive but for the $M_{Z'}=$4~TeV case the extra observables 
could be important for disentangling the models.

\section{CONCLUSIONS}

In this contribution we examined the potential of the ILC to 
distinguish between different models that predict $Z'$ bosons.  
What we found is that it is an extremely powerful tool and would be 
crucial for disentangling this sort of physics if a discovery were 
made at the LHC.  In previous work that concentrated on leptonic 
couplings there were ambiguities.  If the ILC detectors have 
reasonable $b$ and $c$-quark tagging efficiencies  
additional useful information could be obtained.  We also demonstrated the 
importance of polarization.  In this report we touched upon the 
couplings of variations of the Little Higgs models.  A more detailed 
account of this aspect of our work will be given elsewhere.

\begin{acknowledgments}
This work was supported in part by the Natural Sciences and 
Engineering Research Council of Canada. 
\end{acknowledgments}

%\begin{thebibliography}{9}   % Use for  1-9  references


\begin{thebibliography}{99} % Use for 10-99 references


%\cite{Leike:1998wr}
\bibitem{Leike:1998wr}
  A.~Leike,
  %``The phenomenology of extra neutral gauge bosons,''
  Phys.\ Rept.\  {\bf 317}, 143 (1999)
  [arXiv:hep-ph/9805494].
  %%CITATION = HEP-PH 9805494;%%

%\cite{Cvetic:1995zs}
\bibitem{Cvetic:1995zs}
  M.~Cvetic and S.~Godfrey,
  %``Discovery and identification of extra gauge bosons,''
  arXiv:hep-ph/9504216.
  %%CITATION = HEP-PH 9504216;%%

%\cite{Aguilar-Saavedra:2001rg}
\bibitem{Aguilar-Saavedra:2001rg}
  J.~A.~Aguilar-Saavedra {\it et al.}  [ECFA/DESY LC Physics Working Group],
  %``TESLA Technical Design Report Part III: Physics at an e+e- Linear
  %Collider,''
  hep-ph/0106315.
  %%CITATION = HEP-PH 0106315;%%

\bibitem{Weiglein:2004hn}
  G.~Weiglein {\it et al.}  [LHC/LC Study Group],
  %``Physics interplay of the LHC and the ILC,''
  hep-ph/0410364.
  %%CITATION = HEP-PH 0410364;%%

%\cite{Godfrey:1994qk}
\bibitem{Godfrey:1994qk}
  S.~Godfrey,
  %``Comparison of discovery limits for extra Z bosons at future colliders,''
  Phys.\ Rev.\ D {\bf 51}, 1402 (1995)
  [arXiv:hep-ph/9411237].
  %%CITATION = HEP-PH 9411237;%%

%\cite{Capstick:1987uc}
\bibitem{Capstick:1987uc}
  S.~Capstick and S.~Godfrey,
  %``A Comparison Of Discovery Limits For Extra E(6) Neutral Gauge Bosons At
  %Future Colliders,''
  Phys.\ Rev.\ D {\bf 37}, 2466 (1988).
  %%CITATION = PHRVA,D37,2466;%%

\bibitem{Rizzo:2004kr} Recent pedagogical introductions to extra 
dimensions are given by
  T.~G.~Rizzo,
  %``Pedagogical introduction to extra dimensions,''
  eConf {\bf C040802}, L013 (2004)
  [hep-ph/0409309]
  %%CITATION = HEP-PH 0409309;%%
and
%\cite{Cheung:2004ab}
%\bibitem{Cheung:2004ab}
  K.~Cheung,
  %``Collider phenomenology for a few models of extra dimensions,''
  hep-ph/0409028.
  %%CITATION = HEP-PH 0409028;%%

%\cite{Arkani-Hamed:2002qy}
\bibitem{Arkani-Hamed:2002qy}
  N.~Arkani-Hamed, A.~G.~Cohen, E.~Katz and A.~E.~Nelson,
  %``The littlest Higgs,''
  JHEP {\bf 0207}, 034 (2002)
  [arXiv:hep-ph/0206021].
  %%CITATION = HEP-PH 0206021;%%

%\cite{Schmaltz:2004de}
\bibitem{Schmaltz:2004de}
  M.~Schmaltz,
  %``The simplest little Higgs,''
  JHEP {\bf 0408}, 056 (2004)
  [arXiv:hep-ph/0407143].
  %%CITATION = HEP-PH 0407143;%%

%\cite{Schmaltz:2005ky}
\bibitem{Schmaltz:2005ky}
  M.~Schmaltz and D.~Tucker-Smith,
  %``Little Higgs review,''
  arXiv:hep-ph/0502182.
  %%CITATION = HEP-PH 0502182;%%

%\cite{Han:2003wu}
\bibitem{Han:2003wu}
  T.~Han, H.~E.~Logan, B.~McElrath and L.~T.~Wang,
  %``Phenomenology of the little Higgs model,''
  Phys.\ Rev.\ D {\bf 67}, 095004 (2003)
  [arXiv:hep-ph/0301040].
  %%CITATION = HEP-PH 0301040;%%

%\cite{Han:2005ru}
\bibitem{Han:2005ru}
  T.~Han, H.~E.~Logan and L.~T.~Wang,
  %``Smoking-gun signatures of little Higgs models,''
  arXiv:hep-ph/0506313.
  %%CITATION = HEP-PH 0506313;%%

\bibitem{godfrey}
Preliminary results were given in S. Godfrey, contribution to  
{\sl Physics interplay of the LHC and the ILC} (unpublished), 
see \cite{Weiglein:2004hn};
S. Godfrey, P. Kalyniak, A. Tomkins, in preparation.

\bibitem{e+e-} For some early references see:
F. Boudjema, B.W. Lynn, F.M. Renard, C. Verzegnassi, 
Z. Phys. {\bf C48}, 595 (1990);
A. Blondel, F.M. Renard, P. Taxil, and C. Verzegnassi, Nucl. Phys. 
{\bf B331}, 293 (1990);
%\bibitem{Belanger:1986xx}
  G.~Belanger and S.~Godfrey,
  %``Asymmetries At E+ E- Colliders Form E(6) Grand Unified Theories,''
  Phys.\ Rev.\ D {\bf 34}, 1309 (1986);
  %%CITATION = PHRVA,D34,1309;%%
%\bibitem{Belanger:1986tv}
%  G.~Belanger and S.~Godfrey,
  %``Asymmetries At E+ E- Colliders From E(6) Grand Unified Theories,''
  Phys.\ Rev.\ D {\bf 35}, 378 (1987).
  %%CITATION = PHRVA,D35,378;%%
P.J. Franzini and F.J. Gilman, Phys. Rev. {\bf D35}, 855 (1987);
M. Cveti\v{c} and B. Lynn, Phys. Rev. {\bf D35}, 1 (1987);
B.W. Lynn and C. Verzegnassi, Phys. Rev. {\bf D35}, 3326 (1987);
T.G. Rizzo {\it ibid}, {\bf 36}, 713(1987);
A. Bagneid, T.K. Kuo, and G.T. Park {\it ibid}, {\bf 44}, 2188 (1991);
A. Djouadi {\it et al}, 
%A. Leike, T. Riemann, D. Schaile and C. Verzegnassi, 
Z. Phys. {\bf C56} 289 (1992); 
A. Leike, Z. Phys. {\bf C62}, 265 (1994);



\bibitem{Moortgat-Pick:2005cw}
  G.~Moortgat-Pick {\it et al.},
  %``The role of polarized positrons and electrons in revealing fundamental
  %interactions at the linear collider,''
  arXiv:hep-ph/0507011.
  %%CITATION = HEP-PH 0507011;%%

%\cite{Fujii:1995ys}
\bibitem{Fujii:1995ys}
  K.~Fujii and T.~Omori,
  %``Impact of double beam polarization on sin**2 theta-w determination,''
KEK-PREPRINT-95-127



\end{thebibliography}
\end{document}